# Mapping molecular complexes with Super-Resolution Microscopy and Single-Particle Analysis

Afonso Mendes[1,*], Hannah S. Heil[1,*], Simao Coelho[1], Christophe Leterrier[2], and Ricardo Henriques[1,3,✉]

[1] Instituto Gulbenkian de Ciência, Oeiras, Portugal
[2] Aix Marseille Université, CNSR, INP UMR7051, NeuroCyto, Marseille, France
[3] MRC Laboratory for Molecular Cell Biology, University College London, London, United Kingdom
[*] These authors contributed equally

Understanding the structure of supramolecular complexes provides insight into their functional capabilities and how they can be modulated in the context of disease. Super-resolution microscopy (SRM) excels in performing this task by resolving ultrastructural details at the nanoscale with molecular specificity. However, technical limitations, such as underlabelling, preclude its ability to provide complete structures. Single-particle analysis (SPA) overcomes this limitation by combining information from multiple images of identical structures and producing an averaged model, effectively enhancing the resolution and coverage of image reconstructions. This review highlights important studies using SRM-SPA, demonstrating how it broadens our knowledge by elucidating features of key biological structures with unprecedented detail.

**structural biology** | **super-resolution microscopy** | **single-particle analysis**
Correspondence: *rjhenriques@igc.gulbenkian.pt*

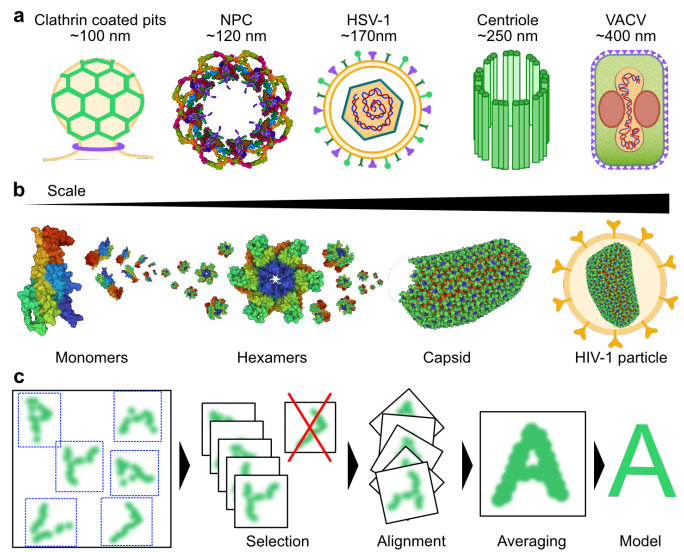

**Fig. 1. Assembly of supramolecular complexes. (a)** Supramolecular complexes such as clathrin-coated pits, the nuclear pore complex (NPC), the Herpes simplex virus type 1 (HSV-1), the centriole and the Vaccinia virus (VACV) comprise highly ordered structures across a size range of several hundred nanometers. **(b)** Structural redundancy occurs at different scales within the same biological assembly. For example, the Human Immunodeficiency Virus type 1 (HIV-1) capsid is composed of at least 1500 copies of the HIV-1 capsid protein (CA) (2), which assemble into approximately 250 hexameric CA complexes before congregating to form the fully-assembled capsid (3, 4). **(c)** Supramolecular complexes can be mapped by combining SRM with SPA. For this, several image sections containing super-resolved views of single particles are chosen. A filtering step removes non-matching objects based on predefined criteria. The remaining image sections are aligned and an averaged view of the object's architecture is generated, which is then used to infer a model of the supramolecular structure. (NPC and HIV-1 CA structures were created using Mol* (6); PDB IDs: 7N9F (NPC), 2M8N (CA monomer), 6OBH (CA hexamer), and 3J3Y (HIV-1 capsid)).

## Introduction

Organized assemblies containing several copies of one or more molecular entities are a hallmark of the living world and their existence underlies every biological process. Studying their structure and assembly dynamics is crucial to understanding their formation, function, and resulting activity. Typically, these structural assemblies are characterised by specific molecular interactions, recruitment, conformational changes, and catalysis of subsequent reactions. These complex supramolecular constructs range from cellular components to viruses and display structural redundancy across different levels of complexity (Fig. 1a). For example, the Human Immunodeficiency Virus type 1 (HIV-1) capsid, which houses and directs the intracellular trafficking of the viral genome towards the nucleus during infection (1), is a ~40 MDa supramolecular complex containing at least 1500 copies of the HIV-1 capsid protein (CA) (2). Notably, its formation involves assembling the CA monomers into approximately 12 pentameric and 250 hexameric complexes and their subsequent congregation to form a final and more complex structure (3, 4) (Fig. 1b).

Among the experimental tools to study biological structures, microscopy stands out by allowing their direct observation. In particular, fluorescence microscopy allows observations with molecular specificity. The spatial resolution achievable with conventional light-based microscopy is limited to approximately half of the wavelength (~200 nm) (5). Electron microscopy (EM) can resolve structures well below the diffraction limit but provides limited information on the identity of molecular components and the precise location of a substantial population of molecules within very small complexes. The advent of super-resolution microscopy (SRM) has enabled nano-sized structures to be resolved while retaining the molecular contrast provided by molecule-specific fluorescent labelling.

Despite the potential of SRM to resolve small molecular structures, factors such as low labelling densities and local-isation uncertainty hinder its ability to map supramolecular complexes with high precision and generate complete models. Single-Particle Analysis (SPA), sometimes called "particle averaging", is an analytical method that combines information from several views of a structure of interest, termed particle, and produces an averaged model (Fig. 1c). It was first developed in the field of cryo-EM (7–10) but has recently been applied successfully in SRM. Despite fundamental differences between EM and SRM, SPA can overcome important SRM limitations, such as under-labelling due to stochastic fluorophore excitation. SPA is primarily used to map biological structures by generating image reconstructions with a higher signal-to-noise ratio (SNR) than the original images used.

Here, we provide an overview of SPA applications in SRM, discuss current limitations, and consider future developments.

## SPA overcomes critical SMLM limitations

Current SRM technologies can be classified into two broad categories based on the principle underlying their ability to achieve sub-diffraction resolutions. One type includes methods that modulate the light's path by exploring different illumination schemes, such as Structured Illumination Microscopy (SIM) (11) and Stimulated Emission Depletion (STED) microscopy (12). The second category includes modalities that exploit the intrinsic properties of fluorophores, namely the ability to control their stochastic "on/off" switching (i.e., "blinking") and their excitation/emission spectra (i.e., photoconversion). This last category contains methods such as Photoactivated Localisation Microscopy (PALM) (13), direct Stochastic Optical Reconstruction Microscopy (dSTORM) (14, 15) and Point Accumulation for Imaging in Nanoscale Topology (PAINT) (16, 17), which enable the reconstruction imaged structures from the localisation coordinates of single and isolated fluorophores. These are known as Single Molecule Localisation Microscopy (SMLM) modalities. At the interface of these two categories lies "MINimal photon FLUXes" (MINFLUX), which uses a doughnut-shaped excitation light beam to detect the presence of a fluorophore and then triangulates its exact localisation by refining the beam's position until the doughnut's zero centre matches the fluorophore's position, thus, minimizing the flux of photons (18). A more detailed explanation of the principles, advantages, and disadvantages of each SRM modality is available in (19).

Despite its ability to reveal the localisation of individual molecules within supramolecular complexes, SMLM has notable limitations such as labelling artifacts, target-to-label distances (i.e., linking error), and lack of temporal information. For example, single proteins are typically imaged using antibody-based labelling (i.e., immunolabelling) or fusions between fluorescent tags and the proteins of interest. In immunolabelling, the antibody's epitope is separated from the fluorophore by a short linker, resulting in a slight displacement of the fluorescent signal relative to the actual binding site in the target molecule. In turn, fusions with fluorescent tags minimise this linking error. Still, they can disturb the target proteins' function, an effect often related to steric hindrance or alterations in their spatial distribution and binding affinities (e.g., (20)). Another caveat of SMLM is related to the density and coverage of the labelling. SMLM stochastically samples the distribution of labels in a specimen (15). Thus, there is no guarantee that all the fluorophores in a sample are acquired or excited proportionally. Furthermore, in immunolabeling, different regions within a supramolecular complex might not be equally accessible to the labelling antibodies, resulting in heterogeneous or incomplete coverage. In addition, immunolabeling is performed using chemically fixed structures. Although this allows for a snapshot of critical cellular activities and subsequent nanometer characterisation, the spatial-temporal progression of the cellular activities is confined to the time of fixation and may contain fixation-induced artifacts.

SPA is a computational approach that significantly enhances structural studies using microscopy. In SPA, particles are compared, and goodness-of-fit metrics such as cross-correlation (21, 22) evaluate the particle before averaging. The typical procedure of SPA comprises 1) image acquisition, 2) particle detection and segmentation, 3) selection, 4) alignment, 5) averaging, and 6) model generation. An example pipeline showcasing a dSTORM dataset containing the nuclear envelope of a Xenopus oocyte labelled for the NPC protein gp210 (23) (data available at doi:10.5281/zenodo.506824) is shown in Fig. 2.

## Mapping supramolecular structures at the nanoscale with SRM-SPA

The studies referenced in the following sections highlight the potential and limitations of SRM and SPA in mapping the architecture of biological molecular assemblies.

**The Nuclear Pore Complex.** The NPC mediates and regulates the translocation of molecules across the nuclear envelope. It is one of the largest supramolecular complexes in the eukaryotic cell, with a diameter of ~125 nm and a height of ~70 nm. Its cylindrical structure with eightfold symmetry is composed of multiple copies of at least 30 different proteins called nucleoporins (Nups) that form a cytoplasmic ring, a nuclear ring, and a central channel with a diameter of 35-50 nm (25–27) (Fig. 1a). Molecules crossing via the NPC need to be at least slightly smaller than its central channel. Thus, knowing the exact diameter of the NPC is crucial to understand which molecules or molecular complexes can travel across it. This becomes particularly relevant in the context of infection with pathogens trying to reach the host's genome, such as HIV-1 (28). In addition, owing to key structural features such as large size, radial symmetry and low variability between individual entities, NPCs became a preferential biological structure to evaluate SRM-SPA methods (29). Furthermore, while averaged reconstructions of NPC densities were made using EM-SPA (30), the organisation of its components within the complex remained unknown.

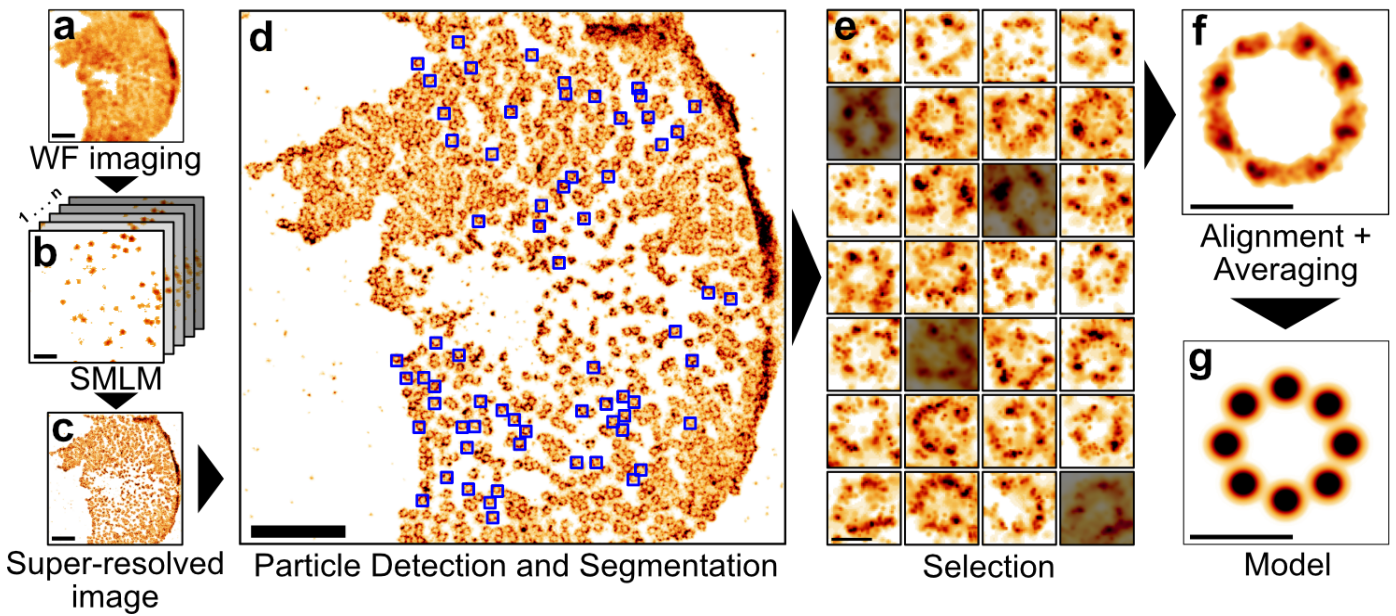

**Fig. 2. Example SPA framework using dSTORM data of the NPC protein gp210. (a-c)** A super-resolved image is acquired before the analysis. **(d)** Particles corresponding to the structure of interest are manually or automatically detected from a super-resolved image, generating an image library containing several segmented particles. **(e)** The particle population is filtered to remove unwanted objects according to specific criteria. **(f)** Each particle is aligned and used in an iterative averaging process (24), resulting in a final model **(g)** with enhanced structural accuracy. Scale bars represent 2 μm (a-d) and 100 nm (e-g).

Schermelleh et al. (31) used 3D SIM to resolve single NPCs with molecular specificity, but the spatial resolution was insufficient to reveal fine structural details. Loschberger et al. (32) tackled this problem using dSTORM to image NPCs in nuclear envelopes isolated from Xenopus laevis (frog) oocytes. They focused on gp210, a protein that forms homodimers and then multimers that surround and anchor the NPC on the luminal side of the nuclear pore membrane (33). Previous experiments using biochemical tools suggested a model where at least eight gp210 dimers were present in each NPC (34). Loschberger et al. (Löschberger et al., 2012) (32) resolved an eightfold symmetrical arrangement of gp210 and calculated an average diameter for the complex of 161 ± 17 nm. They generated images from 426 individual rings (~160 000 localisations) that were automatically detected and aligned to confirm the statistically observed eightfold symmetry. The final averaged image displayed the eightfold symmetry of the gp210 ring previously observed and a diameter of 164 ± 7 nm. Attempts to distinguish gp210 monomers and dimers failed because the linkage error (~18 nm) was too high to achieve the required spatial resolution. Finally, the authors exploited the high binding affinity of fluorescently labelled wheat germ agglutinin for N-acetylglucosamine-modified Nups to resolve the inner lining of the central channel of the NPC. The same analytical approach was performed by combining 621 rings (~40 000 localisations) and calculating a 41 ± 7 nm pore diameter.

Szymborska et al. (35) extended this type of approach to other components of the NPC using a distinct SPA routine. They labelled Nup133 in whole cells and mapped the position of the fluorophores relative to the centre of the nuclear pore with a precision of 0.1 nm and an accuracy of 0.3 nm. The authors conducted a thorough characterisation of the NPC's structure by labelling several members of the Nup107-160 complex, a primary component of the NPC. In doing so, they addressed the existence of three contradictory models for the orientation of the subcomplex and the radial position of its components on the nanometer scale (36–38). The same SPA framework was used to map the relative positions of the proteins in the complex. Nup-GFP fusion proteins and dye-coupled anti-GFP nanobodies were used to minimise the linking error and overcome the inability to obtain antibodies for several Nup107-160 members working in SRM. Using this methodology, they calculated similar ring diameters, and the differences between the two approaches corresponded to the difference in the linker's size between the two approaches. These measurements concluded that a head-to-tail arrangement of the Nup107-160 complexes along the nuclear pore's circumference most likely explained the data. Importantly, they showed that the developed SPA implementation could be applied to asymmetric structures without losing precision if a molecular reference in a second colour channel was used.

Many studies on the nanoarchitecture of NPCs followed, and this subject remains an intense area of research today. Some of these studies will be discussed further in light of the SPA implementations used (e.g., LocMoFit (39), Fig. 3a).

**Viruses.** Viruses are among the smallest biological structures known. Due to their small size, viral structures have historically been described using EM (40). However, this approach is limited to mapping specific molecules' position and molecular identity within viral supramolecular complexes. SRM-SPA overcomes this limitation by using data from thousands of labelled viral particles to provide averaged 3D models with molecular specificity.

Moreover, Gray et al. (41) developed VirusMapper. This SPA framework automatically performs detection, segmentation, alignment, classification, and averaging thousands of individual viral particles to generate high-resolution reconstructions of viral particles with substructural detail. Importantly, VirusMapper is a user-friendly and open-source ImageJ/Fiji implementation. The authors evaluated the analytic capabilities of the framework using multi-colour SIM and STED images of Vaccinia virus (VACV), the smallpox vaccine virus (42). VirusMapper enabled mapping the relative localisations of several viral proteins within the fully assembled virion. Importantly, it also allowed the detection of changes in the arrangement of a particular viral protein (L4) in the viral complex induced by the virion's fusion with the cell membrane, a feature that even EM had not been able to demonstrate robustly. In a follow-up study (43), Gray et al. used VirusMapper and an extensive collection of virus mutants to investigate the nanoscale organisation of the VACV entry complex, a supramolecular complex located on the viral membrane that is crucial for binding and fusion with the membrane of the host cell. They determined that fusion activity and success correlated with a specific orientation of the virion and the spatial distribution of viral binding and fusion proteins (Fig. 3b).

The Herpes simplex virus type 1 (HSV-1) is a common cause of cold sores but can also cause blindness and encephalitis (44, 45). It is highly contagious and establishes a remarkably persistent and latent infection in sensory ganglia with periodic reactivation leading to symptomatic or asymptomatic virus shedding (46). The structure of the virus, a ~200 nm diameter sphere with a 125 nm diameter icosahedral capsid housing the viral DNA genome, was first determined by EM/ET (47–49). The most complex layer of the virus is called tegument and contains over 20 proteins (50). Laine et al. (51) used two-colour dSTORM to map the relative localisation of several HSV-1 proteins in the viral envelope and tegument. Using viral proteins fused with fluorescent proteins, they resolved individual capsids inside infected cells. They could discriminate between enveloped and non-enveloped particles and reveal the subcellular localisation of viral components and their relative amounts in the cytoplasm and the nucleus. With this information, the authors classified the structures resolved based on the presence of a capsid, tegument, or envelope, effectively achieving nanoarchitecture-based characterisation of single virions. A SPA framework was developed and used to map more than 50 viral particles. The algorithm generates averaged models for individual viral proteins based on their localisation and radii distribution. Also, the thickness of each protein layer is estimated from the population's diameter variability and deviation from spherical symmetry (Fig. 3c).

Human immunodeficiency virus type 1 (HIV-1) is the causative agent of acquired immunodeficiency syndrome (AIDS). The HIV-1 genome encodes at least 12 functional proteins (52), but several cellular factors are required for its successful replication, including members of the endosomal sorting complexes required for transport (ESCRT) (53–55). Understanding the nanoscale organisation and function of ESCRT subcomplexes at viral assembly sites is crucial to elucidate the mechanisms underlying HIV propagation. Van Engelenburg et al. (56) used PALM to image ESCRT subcomplexes that interact directly with HIV-1 Gag (the main component of the HIV-1 capsid) and play a role in HIV-1 membrane abscission. They expressed ESCRT proteins fused with fluorescent proteins in the COS7 cell line and resolved structural details of viral assembly sites at the plasma membrane. Furthermore, using a Gag-mEOS2 protein fusion as a reference, they detected ESCRT subcomplexes forming clusters at the plasma membrane that were discernible from their corresponding cytosolic pools. Importantly, a two-colour SPA approach allowed the authors to quantitatively determine that large fractions of some ESCRT proteins become trapped inside the HIV-1 Gag lattice of viral-like particles during viral assembly. Furthermore, the deletion of a specific amino acid motif in Gag precluded this phenomenon.

**Clathrin-mediated endocytosis.** Clathrin-mediated endocytosis (CME) is critical for many biological processes, such as signalling, nutrient uptake, and pathogen entry. CME involves the assembly of supramolecular complexes containing clathrin and many other proteins at the plasma membrane, leading to the inward budding of an intracellular vesicle. The vesicle is then pinched off the plasma membrane and rapidly uncoats to allow its fusion with endosomes. Because of the high conservation of CMEs between yeast and humans, yeast models are often used to study their structures and dynamics. The endocytic machinery comprises subdiffraction limit structures extensively characterised using microscopy and SPA.

Berro and Pollard (57) developed a SPA framework that tracks protein patches located at the endocytic sites and then performs particle alignment and averaging. This approach captures in super-resolution the endocytic pathway as a function of the conformational state. This is done by classifying different conformational states within the membrane, thus inferring sequential progress. Among other discoveries, the authors mapped the average positions of endocytic proteins along membrane invagination using confocal fluorescence microscopy. Furthermore, Picco et al. (58) developed a similar SPA approach to characterise protein dynamics during vesicle budding. They combined live-cell imaging data with previous EM structural data and characterised the endocytic process thoroughly. Finally, Mund et al. (59) developed a SPA framework that uses SRM images to characterise the dynamics of vesicle budding during CME. They fluorescently tagged 23 different endocytic proteins and imaged thousands of fixed yeast cells, collecting data from more than 100,000 endocytic sites. Using live cell imaging data as a temporal reference, the average positions of labelled proteins at endocytic sites were mapped with unprecedented accuracy and 3D models of their distribution at different stages of endocytosis were generated (Fig. 3d).

**Cilia.** Cilia are small organelles that take the shape of a protrusion projecting from the cell body. Their type functions as cellular antennas to integrate environmental signals or create extracellular flows by continuous beating. They are generated from the basal body, a cylindrical structure composed of triplet microtubules arranged with ninefold symmetry (62–65). The skeleton of cilia, called the axoneme (66), comprises microtubule doublets arranged with ninefold symmetry around a central microtubule doublet and is contained by a membrane contiguous to the plasma membrane. In humans, the importance of cilia is highlighted by several diseases, called ciliopathies, which arise from defective ciliary function typically caused by mutations. These include polycystic kidney disease (PKD), polydactyly, and Joubert syndrome (JBTS).

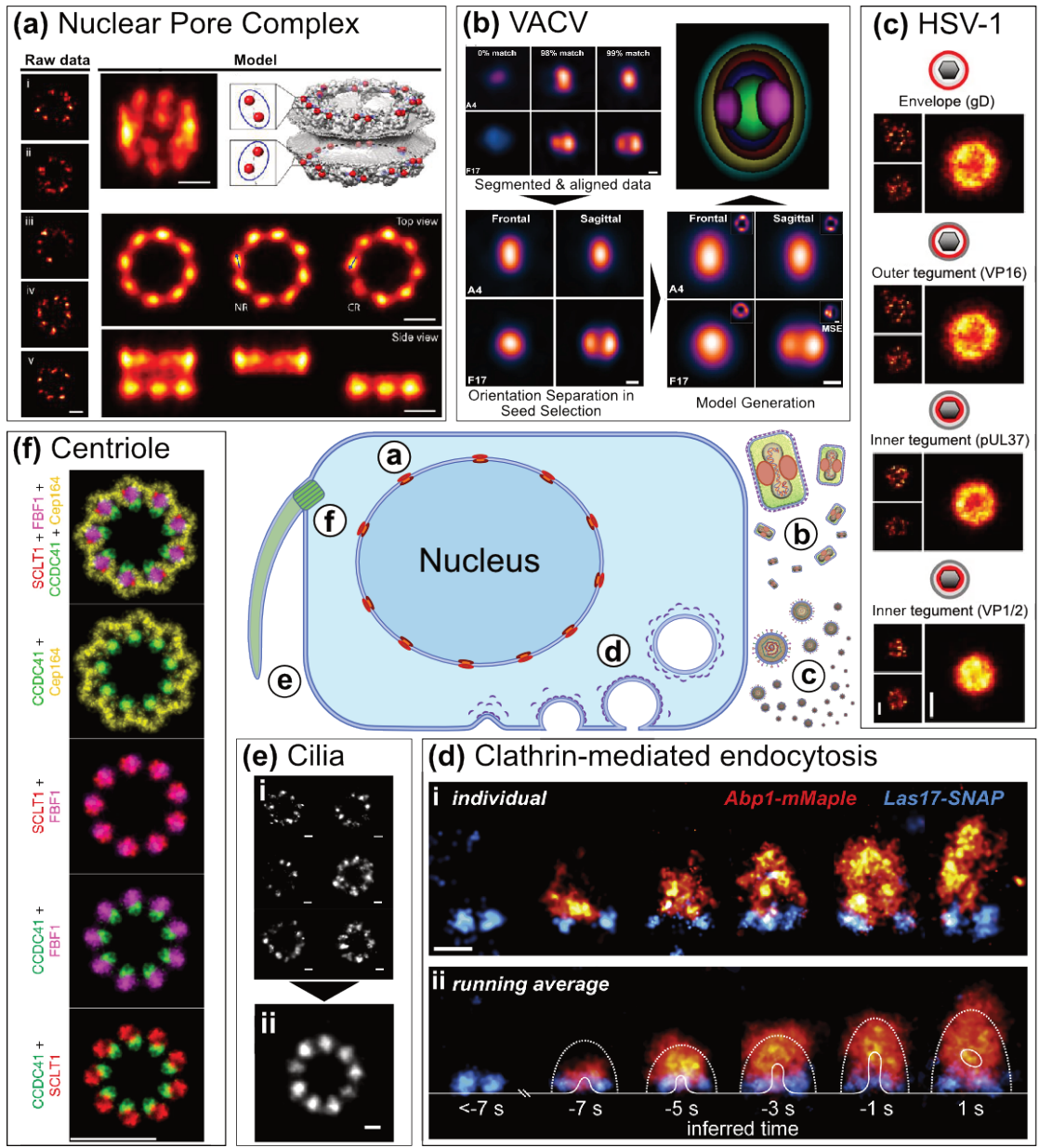

**Fig. 3. Overview of biological structures analysed with SRM-SPA. (a) Nuclear Pore Complex.** Single NPC particles (raw data) were assumed to represent samples of the same underlying distribution. A particle was found (in the example the particle iv) so that it best describes all other sites based on the rank on sum log-likelihood (LL) of the all-to-all matrix, where the 50 subset sites were fitted to each other. The initial template is built based on sequential registration in the order of the sum LL rank. The final fused particle is used to register all sites in the 318-site data set. This procedure yields an updated fused particle, which is used to register the data set again. This process is iterated until convergence. The final average (model) was calculated from 318 particles without any assumption on the underlying geometry or symmetry in a tilted view (mode, top left). For comparison, the EM density of the NPC with C-termini of Nup96 is indicated in red (model, top right). Top and side views (model, bottom), where the nucleoplasmic and cytoplasmic rings are shown together (left panel), or separately (middle, right panels) (vii) The two proteins per ring per symmetric unit give rise to tilted elongated average protein distributions in the averages (arrows in model, bottom). Scale bars represent 50nm. Adapted from (39). **(b) Vaccinia virus.** Segmented and aligned particles are used to generate multi-component models of single virions. For this, more than one seed selection criterion is applied to a single reference channel (VACV A4 frontal and sagittal pictured). Additional virion components (VACV F17 pictured) are aligned to the reference for generating virion orientation-based (frontal and sagittal) models of various viral components. A composite image of all sagital models is shown (A17(cyan)/ CM(yellow)/ A4(red)/ DNA(blue)/ L4(green)/ F17(magenta)). Scale bar represents 100 nm. Adapted from (41). **(c) Herpes simplex virus type 1.** Virus images obtained from aligned particles (right, larger insets) and individual representative particles (left, smaller insets). Scale bar represents 100 nm. Adapted from (51). **(d) Clathrin-mediated endocytosis.** (i) Dual-color side-view super-resolution images of Las17-SNAP and Abp1-mMaple at individual sites. Images were rotated so endocytosis occurs upward and sorted by the distance of Abp1 centroid to Las17 at the base. (ii) Averages of Las17 and Abp1 at endocytic sites. For comparison, average outer boundaries of the actin network (dotted lines), and average plasma membrane profiles (solid line) obtained by correlative light-electron microscopy (60) are overlaid for each time point, as inferred from the images. Scale bar represents 100 nm. Adapted from (59). **(e) Cilia.** (i) 2D STORM images of the ciliary distal appendages of mTEC cells with CEP164 labeled. The STORM localisations of an individual structure are fitted to an ellipse, which is then deformed to a circle. The circularised structure is normalised to a ring with a fixed diameter calculated by averaging the diameter of 31 original structures. Image (ii) shows the resulting averaged structure after alignment. Scale bar represents 100 nm. Adapted from (24). **(f) Centriole.** RPE-1 C1-GFP cells were immunolabeled for the indicated proteins and imaged first in a wide-field mode, followed by 3D STORM imaging, and correlative electron microscopy analysis. Averaged STORM signals were pseudocoloured, rotated and then superimposed to generate a horizontal distributional map of the DAPs. Scale bar represents 200 nm. Adapted from (61).

Similar to NPCs, cilia are the subject of several SRM studies. In particular, the "transition zone" separates the ciliary and plasma membranes and is proposed to function as a gate to the cilium (67). EM studies on the transition zone's structure revealed "Y"-shaped densities called Y-links, with the stem anchored at the microtubule doublets and the two arms attached to the ciliary membrane (68). However, the Y-links' components and the arrangement of proteins in the transition zone were unknown. Shi et al. (69) imaged the transition zone using a two-colour 3D STORM. They observed several proteins forming rings with different diameters localised between the axoneme and the ciliary membrane. Based on the radial, angular, and axial distribution of the ciliary components, they constructed a 3D map of the transition zone with a resolution of 15-30 nm, demonstrating that the protein rings observed were consistent with being Y-links. A protein called Smoothened (SMO) formed a ring of discrete clusters specifically in the transition zone. The authors revealed that certain JBTS-associated mutations reduced SMO localisation in the transition zone and the ciliary membrane.

This study also showed that the diameter of the rings composing the ciliary transition zone, often with an elliptical shape, varied from 369 to 494 nm. This reveals an important structural characteristic of many large cellular organelles called semi-flexibility. Briefly, semi-flexible structures can display slightly different sizes and shapes due to elastic deformation or molecular composition variations while maintaining symmetry and angular arrangement. This heterogeneity poses a problem for particle alignment and averaging, reducing the accuracy of the final reconstructed images. To address this problem, Shi et al. (24) developed SRM-SPA algorithms that take structural flexibility as a degree of freedom for image registration. They work by deforming heterogeneous elliptical structures into more uniform ones and then aligning and averaging the deformed structures. The algorithms were evaluated using simulated and experimental SMLM data of ciliary appendages from mouse tracheal epithelial cells (Fig. 3e). They resolved the ninefold ciliary symmetry and generated multi-colour 3D models with a higher resolution than state-of-the-art algorithms based on rigid structure registration (70–72).

Robichaux et al. (73) combined cryo-ET and STORM imaging to map repeating structures in a specialised structure of the rod sensory cilium called "connecting cilium". Subtomogram averaging provided a structural framework of the connecting cilium, used as a reference to map the distribution of specific molecular components resolved with STORM. This method allowed mapping subcompartments of the connecting cilium with high precision and revealed structural differences in knockouts for basal body proteins.

**Centriole.** The basal ciliary body originates from the centriole, a cellular structure (74) sharing a similar organisation, including microtubule triplets arranged radially with ninefold symmetry (75). They display proximal-distal polarity. The proximal end shows a supramolecular matrix called the pericentriolar material (76–78), which is important for microtubule nucleation and centriole duplication, the latter resulting in a mother and a daughter centriole. The distal end of the mother centriole contains projections called distal appendages (DAs), which are required for ciliogenesis (79–85), microtubule anchoring and centriole positioning (86–88). Several proteins are distributed around the centriole's centre, forming 100-200 nm diameter rings. Due to centrioles' small size and high molecular complexity, mapping the exact distribution of several centriole proteins remains a challenge.

Gartenmann et al. (89) devised a SPA method to resolve the diameter of centriole rings with unprecedented precision in Drosophila larval wing disks. Their framework involved an initial imaging step with 3D SIM to resolve the ring of the well-established centriole protein Asl. Then, several GFP fusions to centriole proteins were imaged with SMLM and using the centre of the Asl rings as initial estimates for the centriole's centroid, localisations within 100 nm were averaged to calculate a new centre. The process was repeated until the centroid positions converged. Finally, the average radius of the rings was calculated, and the distribution of GFP fusion proteins in the centriole was reconstructed. By taking advantage of the high labelling density of 3D SIM, the high precision of SMLM, and the resolution enhancements and statistical robustness of SPA, this framework allowed calculating radii with an accuracy of ± 4-5 nm.

Bowler et al. (61) used a combination of cryo-ET, STORM imaging, and SPA to map the precise location of DA proteins in 3D (Fig. 3f). They revealed the dynamic nature of DAs by demonstrating with ultrastructural detail the reorganisations that occur before and during mitosis.

The centriole is also a central component of the centrosome, a microtubule-organising structure comprising a pair of centrioles responsible for the mitotic spindle formation during cell division. Sieben et al. (72) developed SPARTAN, a graphical user interface employing a novel SRM-SPA framework that generates 3D structure reconstructions from two-channel 2D SMLM data. In this study, centrosomes from a human cell line were purified and concentrated in coverslips by centrifugation before imaging with PALM. A library of densely labelled particles (10%-20% of the entire population) was created, and the particles were aligned and classified according to their orientation. An averaged model for one labelled protein was generated, on which models for other proteins could be mapped. Using this tool, they confirmed the ninefold sym-metry of Cep164, a crucial centriolar component. They also revealed unknown features of the human centriole's architecture, such as closer proximity of the Cep164 N-terminus to the centriolar wall than previously reported (78).

## Quantitative structural descriptions outside the scope of SPA

SPA refers to frameworks that register and average particles to produce a final model with improved SNR. Despite comparable analytical methods (e.g., pair-correlation) sharing key statistical approaches, such as numerical descriptions of ensemble parameters (e.g., particle diameter), these are not well suited to produce improved image reconstructions - unlike SPA.

For example, spatial descriptive statistics such as pair-correlation (90), Ripley's K-function (91) and Density-Based Spatial Clustering of Applications with Noise (92) have been employed to determine clustering states and average cluster sizes of particle populations. Gunzenhäuser et al. (93) developed a quantitative method to reveal the functional and morphological aspects of the HIV-1 assembly based on SMLM data. They analysed hundreds of HIV-1 Gag clusters and determined the number of Gag molecules per cluster, as well as the frequency distribution of the clusters with respect to their radius and aspect ratio. Malkusch et al. (94) evaluated the performance of different cluster analysis methods in the same context. They were able to classify Gag clusters corresponding to three different stages of viral assembly. Finally, Floderer et al. (95) expressed assembly-defective Gag mutants in live cells to follow the trajectories of individual molecules. They calculated the duration and the energy proportions involved in each assembly stage.

A type of analytical tool that has recently been explored in microcopy is deep-learning (DL). Examples of its applications in bioimage analysis are image processing (e.g., denoising) and particle tracking and segmentation, with several user-friendly tools available to researchers as open-source (e.g., (96)). The use of DL in SPA is showcased in (97). Here, six DL models were trained and benchmarked on their ability to recognise the septin ring, a structure arising in cells treated with the actin polymerisation inhibitor cytochalasin D. Models were able to identify hundreds of septin rings with high accuracy. The segmented rings were then used for particle averaging, producing models with sensitivity for slight structural deviations induced by a septin ring-related gene knockout.

Finally, supramolecular complexes can be mapped using different fluorescent labels in the same protein. For example, Mennella et al. (77) resolved kendrin/pericentrin, a component of the pericentriolar material in human centrioles, with antibodies against the protein's N-terminus and a GFP tag in the C-terminus using SIM. The distance from the different fluorescent signal to the center of the structure allowed to determine the protein's orientation in the supramolecular complex, with the C-terminus positioned closer to the centre and the N-terminus extending outwards with radial symmetry. Another example is Leterrier et al. (98), where the nanoscale organisation of the neuronal protein ankG in a region of the axon called the Axon Initial Segment (AIS) was determined using antibodies targeting different domains of the protein. A known spectrin-binding domain of ankG, which is closer to the protein's N-terminus, was colocalized with spectrin bands just under the plasma membrane and displayed a marked periodicism. In contrast, downstream domains closer to the C-terminus progressively lost this periodicity and were found to be located deeper in the cytoplasm than the N-terminus by 3D-STORM.

## Current challenges in SPA

The studies mentioned in the previous sections showcase the potential of SRM-SPA to map the nanoarchitecture details of supramolecular complexes. However, they also highlight current limitations. In general, the increased computation resources and times required to perform the analyses in novel SPA frameworks are minimised by exploiting different mathematical approaches and hardware specifications, such as Fast Fourier Transforms (99) and Graphical Processing Units (e.g., (100)). However, more fundamental challenges remain. In particular, crucial steps in SPA methodologies involve using an initial template. Particles detected in a dataset can be included or excluded from the analysis with respect to how well they match the chosen template. Additionally, particle averaging involves iterative refinement of the shape of an initial template. The template chosen is typically arbitrary, often consisting of a simplified shape recapitulating a priori knowledge on the structures of interest. For example, a simple spherical shape can be used as a template if the structure of interest is a mature HIV-1 or HSV-1 particle. Another option is to select a representative particle from the dataset, which can be a single particle, a subpopulation, or even a class average (e.g., VirusMapper (41)). The problem shared among these approaches is that the final reconstructed structures are biased towards the template used.

Another important limitation of SPA arises from the assumption that all particles represent the same underlying structure. However, this is not necessarily true, as most particle populations are expected to display some degree of heterogeneity. For example, some biological structures sporadically or transiently manifest as variations of their canonical structure, such as the NPC variant with ninefold symmetry (35), semi-flexible ciliary structures (24) and multicore HIV-1 particles (101). These particular structures are underrepresented in the final reconstructions, decreasing their accuracy and remaining undetected.

**Novel particle registration approaches minimise template bias.** Several studies focused on developing so-called "template-free" SPA methodologies, which use non-arbitrary or data-driven templates to overcome template bias. For example, Fortun et al. (102) developed an SPA framework that reconstructs 3D supramolecular assemblies from 2D SRM images of particles in different orientations. While this framework uses a small number of hand-picked templates to detect particles, it estimates their orientations and performs volume reconstructions without resorting to a priori knowledge, creating an initial model that is iteratively refined by averaging. This framework allowed mapping the distribution of Cep63 around the central centriole barrel from STED images with unprecedented accuracy.

Salas et al. (70) used multivariate statistical analysis to classify particles according to their orientation. Representative examples of each class were then selected based on the visual match between the class average and the individual particles and used to perform multireference alignments. The algorithm successfully produced high-resolution 3D reconstructions of DNA origami (103) structures and T4 bacteriophages from SMLM and simulated data.

Furthermore, Heydarian et al. (100) developed an "all-to-all" registration approach. Each particle is registered pairwise against every other particle, producing similarity metrics and estimates of each pair's relative orientation and position. All the possible absolute positions and orientations of each particle are then calculated using a technique from the field of computer vision called "structure from motion" (104). A quality control step is performed to remove outliers and registration errors by inferring the combinations of relative parameters from the absolute parameters obtained previously. These retrodicted parameters are then compared to those found in the all-to-all registration, and pairs deviating from their registered counterparts above a specified threshold are discarded from the next steps. These procedures generate a data-driven template that is further refined by particle averaging. The algorithm's performance was first evaluated using a dataset containing 2D DNA origami particles imaged with DNA-PAINT and generated image reconstructions with ~3 nm resolution. In addition, the algorithm was able to perform using particles with labelling densities as low as 30%. The algorithm was also tested using the NPC dataset in (32) and successfully retrieved the NPC's eightfold symmetry without requiring prior knowledge. More recently, this framework was extended to enable 3D SPA (105).

Another example is Blundell et al. (106), where DL was used to predict each particle's pose. Here, a neural network fits a 3D model against a library of particles viewed in 2D. The framework was able to discern the centriole's expected toroidal structure from the STORM dataset acquired in (72).

It is important to note that, although "template bias" exists when using arbitrary models as an initial reference for SPA, completely discarding prior knowledge on the structures of interest might not be the most sensible action. Indeed, the 'template-free' algorithms showcased so far produce reconstructions that can be further improved by taking prior knowledge into account. On this note, Wu et al. (39) developed LocMoFit. This framework extracts particle geometry from SMLM data and fits the particles to geometric models based on maximum-likelihood estimation (MLE). When the geometry of the structure of interest is known, the user can decide on the class of geometric models to be used for the fitting. Importantly, when an underlying geometry cannot be recovered, the algorithm can perform particle averaging to generate a data-driven template by combining the all-to-all registration methodology developed by Heydarian et al. (100) with MLE. Mund et al. (107) used LocMoFit to extend their work on clathrin-mediated endocytosis developed in (59). Here, they densely labelled clathrin in the human melanoma cell line SK-MEL-2 and used 3D SMLM to resolve the endocytic sites. Then, they performed extensive quantitative descriptions of the structures analysed and were able to infer a spatial-temporal model of the endocytic process.

**Improving sensitivity to detect structure heterogeneity.** Image classification can be used to enhance SPA's lack of sensitivity to particle heterogeneity. This usually involves calculating similarity metrics (e.g., cross-correlation at different relative rotations (21, 22) for particle pairs and then sorting the particles into a suitable number of classes. Several SPA frameworks mentioned so far use classification to group particles according to their viewing orientation (e.g., (41, 72). The particles' relative orientations can be calculated to produce a 3D model from 2D projections with this information.

A more underlooked challenge of SPA that benefits from particle classification is the ability to detect and classify particles representing different underlying structures. This becomes particularly difficult when the various structures are not known a priori and reasonable templates cannot be provided. Huijben et al. (108) extended the pipeline developed in (100) to achieve this goal. In this implementation, the similarity metric obtained from the all-to-all registration strategy is converted into a dissimilarity metric and multidimensional scaling (109) is used to translate the values of each registration pair into spatial coordinates projected in a multidimensional space. The multidimensional particles are then classified using k-means clustering (110) and each class is averaged to generate the final reconstructions. In particular, the authors provide a relatively simple strategy to determine an optimal number of classes without a priori knowledge. The framework was evaluated using a 2D DNA origami test dataset imaged with DNA-PAINT (16, 111), containing particles belonging to several different classes. Under these conditions, the algorithm correctly classified >95% of the analysed particles. The averaged reconstructions achieved resolutions between 3.7 and 5.7 nm. Most importantly, the algorithm correctly classified an NPC with ninefold symmetry in a simulated dataset where this class was highly underrepresented (2% of the total number of particles). Furthermore, the algorithm detected classes corresponding to elliptical NPCs in the Xenopus oocyte NPC dataset mentioned previously (32). Although the presence of these elliptical structures in the data was deemed by the authors to be an artifact generated during sample preparation without any biological significance, their successful detection showcases once again the high sensitivity of the classification framework and its ability to detect unknown structures. In this context, Sabinina et al. (112) used SPA to perform quantitative descriptions and obtain an average model of the NPC. They observed that the nuclear basket protein TPR was distributed over a larger volume than expected, suggesting averaging over different NPC conformations. Accordingly, they found significant variation in descriptive parameters, such as circularity and diameter, which exceeded the expected registration error. Thus, they developed a classification method that does not depend on a priori information about the structures of interest and used it to confirm the presence of different NPC conformations.

## Conclusions and Outlook

Determining the fine structure of supramolecular complexes gives crucial insight into their assembly dynamics and functional capabilities. SRM excels at imaging small biological structures containing multiple components with different molecular identities at high resolutions. It successfully tackles the challenge of surpassing the diffraction limit of light by bringing together cutting-edge microscopy technology, modulation of the physical properties of fluorescent labels, and advanced statistical analysis. Furthermore, the implementation of SPA in SRM substantially improves the accuracy of image reconstructions overcoming crucial caveats of SRM such as under-labelling and labelling heterogeneity. It combines information from thousands of imaged structures and produces reconstructions that effectively contain more localisations than their underlabelled counterparts. This enables the mapping of supramolecular complexes in 3D and retrieving architectural features with single-digit nanometer precision. Additionally, the analysis can be performed in multiple colour channels representing different labelled molecules present in the same molecular assembly, which allows using a reference molecule to register the remaining (e.g., (41)).

Despite the great success of SRM-SPA in mapping supramolecular complexes, its capabilities have not yet been fully explored. So far, SPA studies have focused almost exclusively on generating reconstructions of a unique and fully assembled structure. However, these same structures undergo a process of assembly comprising multiple metastable structures that share crucial molecular players. Resolving the fine structural details of intermediate structures is crucial to describing the assembly dynamics of the corresponding supramolecular complexes. There are several obstacles to mapping metastable structures. For example, the degree to which each structure is represented in the data is a function of its relative stability or frequency, resulting in a heterogeneous representation. Combined with the already existing caveats of SRM (e.g., underlabelling) increases the chances that less stable structures are under-represented, potentially to a point where it becomes hard or even impossible to acquire a sample with enough particles to accurately represent the structure. In addition, structural plasticity as observed in (24) and (112) also contributes to under-representation of a particular structure.

Furthermore, the lack of temporal information due to fixed-sample imaging precludes the mapping of the sequential assembly of supramolecular complexes. This problem was addressed in Berro and Pollard (57), where a 'brute force' approach was taken. Briefly, clathrin-mediated endocytosis was oversampled to ensure that a suitable number of particles representing each metastable structure were detected. Then, the assembly process's temporal hierarchy was estimated from the similarity between metastable structures. A potential solution to this problem might reside in correlating the fine structural details resolved with SRM-SPA with structural and temporal information extracted from live-cell SRM imaging. While still being limited by the constrains of live-cell SRM today, remarkable progress and future development of both fields, SPA and live-cell SRM, will be the key to elucidating the dynamics of the structural rearrangement of molecular complexes within living cells.

**ACKNOWLEDGEMENTS**
We thank Marie-Christine Dabauvalle and Georg Krohne for the nuclear envelope preparation and Xiaoyu Shi for sharing the "Deformed Alignment" code. This work was supported by the Gulbenkian Foundation (A.M., H.S.H., S.C., R.H.) and received funding from the European Research Council (ERC) under the European Union's Horizon 2020 research and innovation programme (grant agreement no. 101001332 to R.H.), the European Molecular Biology Organization (EMBO-2020-IG-4734 to R.H. and ALTF 499-2021 to H.S.H) the Wellcome Trust (203276/Z/16/Z to R.H.), and NHMRC (APP1183588 to S.C.). A.M. would like to acknowledge support from the Integrative Biology and Biomedicine PhD programme from Instituto Gulbenkian de Ciência.

**EXTENDED AUTHOR INFORMATION**
- Afonso Mendes: 0000-0001-7324-555X; afonsomendes92
- Hannah S. Heil: 0000-0003-4279-7022; Hannah_SuperRes
- Simao Coelho: 0000-0002-0763-3172; simaopc
- Christophe Leterrier: 0000-0002-2957-2032; christlet
- Ricardo Henriques: 0000-0002-2043-5234; HenriquesLab

**AUTHOR CONTRIBUTIONS**
A.M. and H.S.H. contributed equally to the preparation of the manuscript and figures. All authors advised in the preparation and edited content.

**COMPETING FINANCIAL INTERESTS**
The authors declare no competing financial interests.

## Bibliography

1. Edward M. Campbell and Thomas J. Hope. HIV-1 capsid: The multifaceted key player in HIV-1 infection. *Nature Reviews Microbiology*, 13:471–483, 7 2015. ISSN 17401534. doi: 10.1038/nrmicro3503.
2. Brady J. Summers, Katherine M. Digianantonio, Sarah S. Smaga, Pei-Tzu Huang, Kaifeng Zhou, Eva E. Gerber, Wei Wang, and Yong Xiong. Modular HIV-1 capsid assemblies reveal diverse host-capsid recognition mechanisms. *Cell Host Microbe*, 26:203–216.e6, 8 2019. ISSN 19313128. doi: 10.1016/j.chom.2019.07.007.
3. Barbie K. Ganser, Su Li, Victor Y. Klishko, John T. Finch, and Wesley I. Sundquist. Assembly and analysis of conical models for the HIV-1 core. *Science*, 283:80–83, 1 1999. ISSN 0036-8075. doi: 10.1126/science.283.5398.80.
4. Su Li, Christopher P. Hill, Wesley I. Sundquist, and John T. Finch. Image reconstructions of helical assemblies of the HIV-1 CA protein. *Nature*, 407:409–413, 2000. doi: 10.1038/35030177.
5. E. Abbe. Beiträge zur theorie des mikroskops und der mikroskopischen wahrnehmung. *Archiv für Mikroskopische Anatomie*, 9:413–468, 12 1873. ISSN 0176-7364. doi: 10.1007/BF02956173.
6. David Sehnal, Sebastian Bittrich, Mandar Deshpande, Radka Svobodová, Karel Berka, Václav Bazgier, Sameer Velankar, Stephen K. Burley, Jaroslav Koča, and Alexander S. Rose. Mol*Viewer: Modern web app for 3D visualization and analysis of large biomolecular structures. *Nucleic Acids Research*, 49:W431–W437, 7 2021. ISSN 13624962. doi: 10.1093/nar/gkab314.
7. Joachim Frank. Single-particle reconstruction of biological macromolecules in electron microscopy - 30 years. *Quarterly Reviews of Biophysics*, 42:139–158, 8 2009. ISSN 00335835. doi: 10.1017/S0033583509990059.
8. Mikhail Kudryashev, Daniel Castaño-Díez, and Henning Stahlberg. Limiting factors in single particle cryo electron tomography. *Computational and Structural Biotechnology Journal*, 1:e201207002, 2012. ISSN 20010370. doi: 10.5936/csbj.201207002.
9. Guang Tang, Liwei Peng, Philip R. Baldwin, Deepinder S. Mann, Wen Jiang, Ian Rees, and Steven J. Ludtke. EMAN2: An extensible image processing suite for electron microscopy. *Journal of Structural Biology*, 157:38–46, 1 2007. ISSN 10478477. doi: 10.1016/j.jsb.2006.05.009.
10. Marin Van Heel, Brent Gowen, Rishi Matadeen, Elena V. Orlova, Robert Finn, Tillmann Pape, Dana Cohen, Holger Stark, Ralf Schmidt, Michael Schatz, and Ardan Patwardhan. Single-particle electron cryo-microscopy: Towards atomic resolution. *Quarterly Reviews of Biophysics*, 33:307–369, 2000. ISSN 00335835. doi: 10.1017/S0033583500003644.
11. M. G. L. Gustafsson. Surpassing the lateral resolution limit by a factor of two using structured illumination microscopy. *Journal of Microscopy*, 198:82–87, 5 2000. ISSN 0022-2720. doi: 10.1046/j.1365-2818.2000.00710.x.
12. T. A. Klar, S. Jakobs, M. Dyba, A. Egner, and S. W. Hell. Fluorescence microscopy with diffraction resolution barrier broken by stimulated emission. *Proceedings of the National Academy of Sciences*, 97:8206–8210, 7 2000. ISSN 0027-8424. doi: 10.1073/pnas.97.15.8206.

13. Eric Betzig, George H. Patterson, Rachid Sougrat, O. Wolf Lindwasser, Scott Olenych, Juan S. Bonifacino, Michael W. Davidson, Jennifer Lippincott-Schwartz, and Harald F. Hess. Imaging intracellular fluorescent proteins at nanometer resolution. *Science*, 313: 1642–1645, 9 2006. doi: 10.1126/science.1127344.
14. Michael J. Rust, Mark Bates, and Xiaowei Zhuang. Sub-diffraction-limit imaging by stochastic optical reconstruction microscopy (STORM). *Nature Methods*, 3:793–795, 10 2006. ISSN 15487091. doi: 10.1038/nmeth929.
15. Sebastian Van De Linde, Anna Löschberger, Teresa Klein, Meike Heidbreder, Steve Wolter, Mike Heilemann, and Markus Sauer. Direct stochastic optical reconstruction microscopy with standard fluorescent probes. *Nature Protocols*, 6:991–1009, 6 2011. ISSN 17542189. doi: 10.1038/nprot.2011.336.
16. Joerg Schnitzbauer, Maximilian T. Strauss, Thomas Schlichthaerle, Florian Schueder, and Ralf Jungmann. Super-resolution microscopy with DNA-PAINT. *Nature Protocols*, 12: 1198–1228, 6 2017. ISSN 17502799. doi: 10.1038/nprot.2017.024.
17. A. Sharonov and R. M. Hochstrasser. Wide-field subdiffraction imaging by accumulated binding of diffusing probes. *Proceedings of the National Academy of Sciences*, 103: 18911–18916, 12 2006. ISSN 0027-8424. doi: 10.1073/pnas.0609643104.
18. Francisco Balzarotti, Yvan Eilers, Klaus C. Gwosch, Arvid H. Gynnå, Volker Westphal, Fernando D. Stefani, Johan Elf, and Stefan W. Hell. Nanometer resolution imaging and tracking of fluorescent molecules with minimal photon fluxes. *Science*, 355:606–612, 2 2017. ISSN 0036-8075. doi: 10.1126/science.aak9913.
19. Guillaume Jacquemet, Alexandre F. Carisey, Hellyeh Hamidi, Ricardo Henriques, and Christophe Leterrier. The cell biologist's guide to super-resolution microscopy. *Journal of Cell Science*, 133, 6 2020. ISSN 14779137. doi: 10.1242/jcs.240713.
20. João I. Mamede, Joseph Griffin, Stéphanie Gambut, and Thomas J. Hope. A new generation of functional tagged proteins for HIV fluorescence imaging. *Viruses*, 13, 3 2021. ISSN 19994915. doi: 10.3390/v13030386.
21. R. Bracewell. Pentagram notation for cross correlation. In McGraw-Hill, editor, *The Fourier Transform and Its Applications*, pages 46–243. New York, 3rd edition edition, 1965.
22. A. Papoulis. In McGraw-Hill, editor, *The Fourier integral and its applications*, pages 244–245. New York, 1962.
23. Hannah S. Heil, Benjamin Schreiber, Ralph Götz, Monika Emmerling, Marie Christine Dabauvalle, Georg Krohne, Sven Höfling, Martin Kamp, Markus Sauer, and Katrin G. Heinze. Sharpening emitter localization in front of a tuned mirror. *Light: Science and Applications*, 7, 12 2018. ISSN 20477538. doi: 10.1038/s41377-018-0104-z.
24. Xiaoyu Shi, Galo Garcia, Yina Wang, Jeremy F. Reiter, and Bo Huang. Deformed alignment of super-resolution images for semi-flexible structures. *PLoS ONE*, 14, 3 2019. ISSN 19326203. doi: 10.1371/journal.pone.0212735.
25. Nadav Elad, Tal Maimon, Daphna Frenkiel-Krispin, Roderick YH Lim, and Ohad Medalia. Structural analysis of the nuclear pore complex by integrated approaches. *Current Opinion in Structural Biology*, 19:226–232, 4 2009. ISSN 0959440X. doi: 10.1016/j.sbi.2009.02.009.
26. Ulrich Scheer, Marie Christine Dabauvalle, Georg Krohne, René Peiman Zahedi, and Albert Sickmann. Nuclear envelopes from amphibian oocytes - From morphology to protein inventory. *European Journal of Cell Biology*, 84:151–162, 3 2005. ISSN 01719335. doi: 10.1016/j.ejcb.2004.12.001.
27. Caterina Strambio-De-Castillia, Mario Niepel, and Michael P. Rout. The nuclear pore complex: Bridging nuclear transport and gene regulation. *Nature Reviews Molecular Cell Biology*, 11:490–501, 7 2010. ISSN 14710072. doi: 10.1038/nrm2928.
28. Vojtech Zila, Erica Margiotta, Beata Turoňová, Thorsten G. Müller, Christian E. Zimmerli, Simone Mattei, Matteo Allegretti, Kathleen Börner, Jona Rada, Barbara Müller, Marina Lusic, Hans Georg Kräusslich, and Martin Beck. Cone-shaped HIV-1 capsids are transported through intact nuclear pores. *Cell*, 184:1032–1046.e18, 2 2021. ISSN 10974172. doi: 10.1016/j.cell.2021.01.025.
29. Jervis Vermal Thevathasan, Maurice Kahnwald, Konstanty Cieśliński, Philipp Hoess, Sudheer Kumar Peneti, Manuel Reitberger, Daniel Heid, Krishna Chaitanya Kasuba, Sarah Janice Hoerner, Yiming Li, Yu Le Wu, Markus Mund, Ulf Matti, Pedro Matos Pereira, Ricardo Henriques, Bianca Nijmeijer, Moritz Kueblbeck, Vilma Jimenez Sabinina, Jan Ellenberg, and Jonas Ries. Nuclear pores as versatile reference standards for quantitative superresolution microscopy. *Nature Methods*, 16:1045–1053, 10 2019. ISSN 15487105. doi: 10.1038/s41592-019-0574-9.
30. Stephen G. Brohawn, James R. Partridge, James R.R. Whittle, and Thomas U. Schwartz. The Nuclear Pore Complex has entered the atomic age. *Structure*, 17:1156–1168, 9 2009. ISSN 09692126. doi: 10.1016/j.str.2009.07.014.
31. Lothar Schermelleh, Peter M Carlton, Sebastian Haase, Lin Shao, Lukman Winoto, Peter Kner, Brian Burke, M Cristina Cardoso, David A Agard, Mats G L Gustafsson, Heinrich Leonhardt, and John W Sedat. Subdiffraction multicolor imaging of the nuclear periphery with 3d structured illumination microscopy. *Science*, 320:1332–1336, 2008.
32. Anna Löschberger, Sebastian van de Linde, Marie Christine Dabauvalle, Bernd Rieger, Mike Heilemann, Georg Krohne, and Markus Sauer. Super-resolution imaging visualizes the eightfold symmetry of gp210 proteins around the nuclear pore complex and resolves the central channel with nanometer resolution. *Journal of Cell Science*, 125:570–575, 2 2012. ISSN 00219533. doi: 10.1242/jcs.098822.
33. Catherine Favreau, Ricardo Bastos, Jean Cartaud, Jean-Claude Courvalin, and Pekka Mustonen. Biochemical characterization of nuclear pore complex protein gp210 oligomers. *European Journal of Biochemistry*, 268:3883–3889, 7 2001. ISSN 00142956. doi: 10.1046/j.1432-1327.2001.02290.x.
34. L Gerace, Y Ottaviano, and C Kondor-Koch. Identification of a major polypeptide of the nuclear pore complex. *Journal of Cell Biology*, 95:826–837, 12 1982. ISSN 0021-9525. doi: 10.1083/jcb.95.3.826.
35. Anna Szymborska, Alex De Marco, Nathalie Daigle, Volker C Cordes, John A G Briggs, and Jan Ellenberg. Nuclear pore scaffold structure analyzed by super-resolution microscopy and particle averaging. *Science*, 341:655–658, 2013.
36. Frank Alber, Svetlana Dokudovskaya, Liesbeth M. Veenhoff, Wenzhu Zhang, Julia Kipper, Damien Devos, Adisetyantari Suprapto, Orit Karni-Schmidt, Rosemary Williams, Brian T. Chait, Andrej Sali, and Michael P. Rout. The molecular architecture of the nuclear pore complex. *Nature*, 450:695–701, 11 2007. ISSN 14764687. doi: 10.1038/nature06405.
37. Stephen G Brohawn, Nina C Leksa, Eric D Spear, Kanagalaghatta R Rajashankar, and Thomas U Schwartz. Structural evidence for common ancestry of the nuclear pore complex and vesicle coats. *Science*, 322:1369–1373, 2008.
38. Kuo Chiang Hsia, Pete Stavropoulos, Günter Blobel, and André Hoelz. Architecture of a coat for the nuclear pore membrane. *Cell*, 131:1313–1326, 12 2007. ISSN 00928674. doi: 10.1016/j.cell.2007.11.038.
39. Yu-Le Wu, Philipp Hoess, Aline Tschanz, Ulf Matti, Markus Mund, and Jonas Ries. Maximum-likelihood model fitting for quantitative analysis of SMLM data. *bioRxiv*, 2021. doi: 10.1101/2021.08.30.456756.
40. Christophe Hourioux, Denys Brand, Pierre-Yves Sizaret, Franck Lemiale, Sarah Lebigot, Francis Barin, and Philippe Roingeard. Identification of the glycoprotein 41 cytoplasmic tail domains of human immunodeficiency virus type 1 that interact with pr55 gag particles. *AIDS Research and Human Retroviruses*, 16:1141–1147, 8 2000. ISSN 0889-2229. doi: 10.1089/088922200414983.
41. Robert D.M. Gray, Corina Beerli, Pedro Matos Pereira, Kathrin Maria Scherer, Jerzy Samolej, Christopher Karl Ernst Bleck, Jason Mercer, and Ricardo Henriques. VirusMapper: Open-source nanoscale mapping of viral architecture through super-resolution microscopy. *Scientific Reports*, 6, 7 2016. ISSN 20452322. doi: 10.1038/srep29132.
42. Lucas Sánchez-Sampedro, Beatriz Perdiguero, Ernesto Mejías-Pérez, Juan García-Arriaza, Mauro Di Pilato, and Mariano Esteban. The evolution of poxvirus vaccines. *Viruses*, 7:1726–1803, 4 2015. ISSN 19994915. doi: 10.3390/v7041726.
43. Robert D.M. Gray, David Albrecht, Corina Beerli, Moona Huttunen, Gary H. Cohen, Ian J. White, Jemima J. Burden, Ricardo Henriques, and Jason Mercer. Nanoscale polarization of the entry fusion complex of vaccinia virus drives efficient fusion. *Nature Microbiology*, 4:1636–1644, 10 2019. ISSN 20585276. doi: 10.1038/s41564-019-0488-4.
44. Asim V. Farooq and Deepak Shukla. Herpes simplex epithelial and stromal keratitis: An epidemiologic update. *Survey of Ophthalmology*, 57:448–462, 9 2012. ISSN 00396257. doi: 10.1016/j.survophthal.2012.01.005.
45. Richard J. Whitley and David W. Kimberlin. Herpes simplex: Encephalitis children and adolescents. *Seminars in Pediatric Infectious Diseases*, 16:17–23, 2005. ISSN 10451870. doi: 10.1053/j.spid.2004.09.007.
46. Michael P. Nicoll, João T. Proença, and Stacey Efstathiou. The molecular basis of herpes simplex virus latency. *FEMS Microbiology Reviews*, 36:684–705, 5 2012. ISSN 01686445. doi: 10.1111/j.1574-6976.2011.00320.x.
47. Jay C. Brown and William W. Newcomb. Herpesvirus capsid assembly: Insights from structural analysis. *Current Opinion in Virology*, 1:142–149, 2011. ISSN 18796265. doi: 10.1016/j.coviro.2011.06.003.
48. Kay Grunewald, Prashant Desai, Dennis C. Winkler, J. Bernard Heymann, David M. Belnap, Wolfgang Baumeister, and Alasdair C. Steven. Three-dimensional structure of herpes simplex virus from cryo-electron tomography. *Science*, 302:1396–1398, 11 2003. ISSN 0036-8075. doi: 10.1126/science.1090284.
49. Z. Hong Zhou, Matthew Dougherty, Joanita Jakana, Jing He, Frazer J. Rixon, and Wah Chiu. Seeing the herpesvirus capsid at 8.5 Å. *Science*, 288:877–880, 5 2000. ISSN 0036-8075. doi: 10.1126/science.288.5467.877.
50. Bernard N. Fields, Diane E. Griffin, David Mahan Knipe, Jane Doe, Robert A. Lamb, and Martin Malcolm A. *Fields Virology*. 4th edtion edition, 2001. ISBN 978-0781718325.
51. Romain F. Laine, Anna Albecka, Sebastian van de Linde, Eric J. Rees, Colin M. Crump, and Clemens F. Kaminski. Structural analysis of herpes simplex virus by optical super-resolution imaging. *Nature Communications*, 6:5980, 5 2015. ISSN 2041-1723. doi: 10.1038/ncomms6980.
52. Rainer Seitz. Human immunodeficiency virus (HIV). *Transfusion Medicine and Hemotherapy*, 43:203–222, 5 2016. ISSN 16603818. doi: 10.1159/000445852.
53. Dimiter G Demirov, Akira Ono, Jan M Orenstein, and Eric O Freed. Overexpression of the N-terminal domain of TSG101 inhibits HIV-1 budding by blocking late domain function. *Proceedings of the National Academy of Science*, 99:955–960, 1 2002.
54. Jennifer E. Garrus, Uta K. von Schwedler, Owen W. Pornillos, Scott G. Morham, Kenton H. Zavitz, Hubert E. Wang, Daniel A. Wettstein, Kirsten M. Stray, Mélanie Côté, Rebecca L. Rich, David G. Myszka, and Wesley I. Sundquist. Tsg101 and the vacuolar protein sorting pathway are essential for hiv-1 budding. *Cell*, 107:55–65, 10 2001. ISSN 00928674. doi: 10.1016/S0092-8674(01)00506-2.
55. Juan Martin-Serrano, Trinity Zang, and Paul D. Bieniasz. HIV-1 and Ebola virus encode small peptide motifs that recruit Tsg101 to sites of particle assembly to facilitate egress. *Nature Medicine*, 7:1313–1319, 12 2001. ISSN 1078-8956. doi: 10.1038/nm1201-1313.
56. Schuyler B. Van Engelenburg, Gleb Shtengel, Prabuddha Sengupta, Kayoko Waki, Michal Jarnik, Sherimay D. Ablan, Eric O. Freed, Harald F. Hess, and Jennifer Lippincott-Schwartz. Distribution of ESCRT machinery at HIV assembly sites reveals virus scaffolding of ESCRT subunits. *Science*, 343:653–656, 2014. ISSN 10959203. doi: 10.1126/science.1247786.
57. Julien Berro and Thomas D. Pollard. Local and global analysis of endocytic patch dynamics in fission yeast using a new "temporal superresolution" realignment method. *Molecular Biology of the Cell*, 25:3501–3514, 11 2014. ISSN 19394586. doi: 10.1091/mbc.E13-01-0004.
58. Andrea Picco, Markus Mund, Jonas Ries, François Nédélec, and Marko Kaksonen. Visualizing the functional architecture of the endocytic machinery. *eLife*, 4, 2 2015. ISSN 2050-084X. doi: 10.7554/eLife.04535.
59. Markus Mund, Johannes Albertus van der Beek, Joran Deschamps, Serge Dmitrieff, Philipp Hoess, Jooske Louise Monster, Andrea Picco, François Nédélec, Marko Kaksonen, and Jonas Ries. Systematic nanoscale analysis of endocytosis links efficient vesicle formation to patterned actin nucleation. *Cell*, 174:884–896.e17, 8 2018. ISSN 10974172. doi: 10.1016/j.cell.2018.06.032.
60. Wanda Kukulski, Martin Schorb, Marko Kaksonen, and John A.G. Briggs. Plasma membrane reshaping during endocytosis is revealed by time-resolved electron tomography. *Cell*, 150:508–520, 8 2012. ISSN 10974172. doi: 10.1016/j.cell.2012.05.046.
61. Mathew Bowler, Dong Kong, Shufeng Sun, Rashmi Nanjundappa, Lauren Evans, Veronica Farmer, Andrew Holland, Moe R. Mahjoub, Haixin Sui, and Jadranka Loncarek. High-resolution characterization of centriole distal appendage morphology and dynamics by

62. Susan K. Dutcher. The awesome power of dikaryons for studying flagella and basal bodies in chlamydomonas reinhardtii. *Cytoskeleton*, 71:79–94, 2 2014. ISSN 19493584. doi: 10.1002/cm.21157.
63. Galo Garcia and Jeremy F. Reiter. A primer on the mouse basal body. *Cilia*, 5, 4 2016. ISSN 20462530. doi: 10.1186/s13630-016-0038-0.
64. Chad G. Pearson and Mark Winey. Basal body assembly in ciliates: The power of numbers. *Traffic*, 10:461–471, 2009. ISSN 13989219. doi: 10.1111/j.1600-0854.2009.00885.x.
65. Anastassiia Vertii, Hui Fang Hung, Heidi Hehnly, and Stephen Doxsey. Human basal body basics. *Cilia*, 5, 3 2016. ISSN 20462530. doi: 10.1186/s13630-016-0030-8.
66. Richard G. W. Anderson. The three-dimensional of the basal body from the rhesus monkey oviduct. *Journal of Cell Biology*, 54:246–265, 8 1972. ISSN 1540-8140. doi: 10.1083/jcb.54.2.246.
67. Francesc R. Garcia-Gonzalo and Jeremy F. Reiter. Scoring a backstage pass: Mechanisms of ciliogenesis and ciliary access. *Journal of Cell Biology*, 197:697–709, 6 2012. ISSN 00219525. doi: 10.1083/jcb.201111146.
68. Norton B. Gilula and Peter Satir. The ciliary necklace. *Journal of Cell Biology*, 53:494–509, 5 1972. ISSN 1540-8140. doi: 10.1083/jcb.53.2.494.
69. Xiaoyu Shi, Galo Garcia, Julie C. Van De Weghe, Ryan McGorty, Gregory J. Pazour, Dan Doherty, Bo Huang, and Jeremy F. Reiter. Super-resolution microscopy reveals that disruption of ciliary transition-zone architecture causes joubert syndrome. *Nature Cell Biology*, 19:1178–1188, 9 2017. ISSN 14764679. doi: 10.1038/ncb3599.
70. Desirée Salas, Antoine Le Gall, Jean Bernard Fiche, Alessandro Valeri, Yonggang Ke, Patrick Bron, Gaetan Bellot, and Marcelo Nollmann. Angular reconstitution-based 3D reconstructions of nanomolecular structures from superresolution light-microscopy images. *Proceedings of the National Academy of Sciences of the United States of America*, 114:9273–9278, 8 2017. ISSN 10916490. doi: 10.1073/pnas.1704908114.
71. Joerg Schnitzbauer, Yina Wang, Shijie Zhao, Matthew Bakalar, Tulip Nuwal, Baohui Chen, and Bo Huang. Correlation analysis framework for localization-based superresolution microscopy. *Proceedings of the National Academy of Sciences of the United States of America*, 115:3219–3224, 3 2018. ISSN 10916490. doi: 10.1073/pnas.1711314115.
72. Christian Sieben, Niccolò Banterle, Kyle M. Douglass, Pierre Gönczy, and Suliana Manley. Multicolor single-particle reconstruction of protein complexes. *Nature Methods*, 15:777–780, 10 2018. ISSN 15487105. doi: 10.1038/s41592-018-0140-x.
73. Michael A. Robichaux, Valencia L. Potter, Zhixian Zhang, Feng He, Jun Liu, Michael F. Schmid, and Theodore G. Wensel. Defining the layers of a sensory cilium with STORM and cryoelectron nanoscopy. *Proceedings of the National Academy of Sciences of the United States of America*, 116:23562–23572, 2019. ISSN 10916490. doi: 10.1073/pnas.1902003116.
74. Michel Bornens. The centrosome in cells and organisms. *Science*, 335:422–426, 1 2012. ISSN 0036-8075. doi: 10.1126/science.1209037.
75. Manuel Bauer, Fabien Cubizolles, Alexander Schmidt, and Erich A Nigg. Quantitative analysis of human centrosome architecture by targeted proteomics and fluorescence imaging. *The EMBO Journal*, 35:2152–2166, 10 2016. ISSN 0261-4189. doi: 10.15252/embj.201694462.
76. Steffen Lawo, Monica Hasegan, Gagan D. Gupta, and Laurence Pelletier. Subdiffraction imaging of centrosomes reveals higher-order organizational features of pericentriolar material. *Nature Cell Biology*, 14:1148–1158, 11 2012. ISSN 14657392. doi: 10.1038/ncb2591.
77. V. Mennella, B. Keszthelyi, K. L. Mcdonald, B. Chhun, F. Kan, G. C. Rogers, B. Huang, and D. A. Agard. Subdiffraction-resolution fluorescence microscopy reveals a domain of the centrosome critical for pericentriolar material organization. *Nature Cell Biology*, 14:1159–1168, 11 2012. ISSN 14657392. doi: 10.1038/ncb2597.
78. Katharina F. Sonnen, Lothar Schermelleh, Heinrich Leonhardt, and Erich A. Nigg. 3D-structured illumination microscopy provides novel insight into architecture of human centrosomes. *Biology Open*, 1:965–976, 10 2012. ISSN 20466390. doi: 10.1242/bio.20122337.
79. Lukáš Čajánek and Erich A. Nigg. Cep164 triggers ciliogenesis by recruiting Tau tubulin kinase 2 to the mother centriole. *Proceedings of the National Academy of Sciences of the United States of America*, 111, 7 2014. ISSN 10916490. doi: 10.1073/pnas.1401777111.
80. Susanne Graser, York Dieter Stierhof, Sébastien B. Lavoie, Oliver S. Gassner, Stefan Lamla, Mikael Le Clech, and Erich A. Nigg. Cep164, a novel centriole appendage protein required for primary cilium formation. *Journal of Cell Biology*, 179:321–330, 10 2007. ISSN 00219525. doi: 10.1083/jcb.200707181.
81. Ichiro Izawa, Hidemasa Goto, Kousuke Kasahara, and Masaki Inagaki. Current topics of functional links between primary cilia and cell cycle. *Cilia*, 4, 12 2015. ISSN 20462530. doi: 10.1186/s13630-015-0021-1.
82. James E. Sillibourne, Ilse Hurbain, Thierry Grand-Perret, Bruno Goud, Phong Tran, and Michel Bornens. Primary ciliogenesis requires the distal appendage component Cep123. *Biology Open*, 2:535–545, 6 2013. ISSN 20466390. doi: 10.1242/bio.20134457.
83. Barbara E. Tanos, Hui Ju Yang, Rajesh Soni, Won Jing Wang, Frank P. Macaluso, John M. Asara, and Meng Fu Bryan Tsou. Centriole distal appendages promote membrane docking, leading to cilia initiation. *Genes and Development*, 27:163–168, 2013. ISSN 08909369. doi: 10.1101/gad.207043.112.
84. Qing Wei, Qingwen Xu, Yuxia Zhang, Yujie Li, Qing Zhang, Zeng Hu, Peter C. Harris, Vicente E. Torres, Kun Ling, and Jinghua Hu. Transition fibre protein FBF1 is required for the ciliary entry of assembled intraflagellar transport complexes. *Nature Communications*, 4, 11 2013. ISSN 20411723. doi: 10.1038/ncomms3750.
85. Xuan Ye, Huiqing Zeng, Gang Ning, Jeremy F. Reiter, and Aimin Liu. C2cd3 is critical for centriolar distal appendage assembly and ciliary vesicle docking in mammals. *Proceedings of the National Academy of Sciences of the United States of America*, 111:2164–2169, 2 2014. ISSN 00278424. doi: 10.1073/pnas.1318737111.
86. Nathalie Delgehyr, James Sillibourne, and Michel Bornens. Microtubule nucleation and anchoring at the centrosome are independent processes linked by ninein function. *Journal of Cell Science*, 118:1565–1575, 4 2005. ISSN 00219533. doi: 10.1242/jcs.02302.
87. Kazuhiro Tateishi, Yuji Yamazaki, Tomoki Nishida, Shin Watanabe, Koshi Kunimoto, Hiroaki Ishikawa, and Sachiko Tsukita. Two appendages homologous between basal bodies and centrioles are formed using distinct odf2 domains. *Journal of Cell Biology*, 203:417–425, 11 2013. ISSN 00219525. doi: 10.1083/jcb.201303071.
88. Gregory Mazo, Nadine Soplop, Won Jing Wang, Kunihiro Uryu, and Meng Fu Bryan Tsou. Spatial control of primary ciliogenesis by subdistal appendages alters sensation-associated properties of cilia. *Developmental Cell*, 39:424–437, 11 2016. ISSN 18781551. doi: 10.1016/j.devcel.2016.10.006.
89. Lisa Gartenmann, Alan Wainman, Maryam Qurashi, Rainer Kaufmann, Sebastian Schubert, Jordan W. Raff, and Ian M. Dobbie. A combined 3D-SIM/SMLM approach allows centriole proteins to be localized with a precision of ~4–5 nm. *Current Biology*, 27:R1054–R1055, 10 2017. ISSN 09609822. doi: 10.1016/j.cub.2017.08.009.
90. Sarah L. Veatch, Pietro Cicuta, Prabuddha Sengupta, Aurelia Honerkamp-Smith, David Holowka, and Barbara Baird. Critical fluctuations in plasma membrane vesicles. *ACS Chemical Biology*, 3:287–293, 5 2008. ISSN 15548929. doi: 10.1021/cb800012x.
91. B. D. Ripley. The second-order analysis of stationary point processes. *Journal of Applied Probability*, 13:255–266, 6 1976. ISSN 0021-9002. doi: 10.2307/3212829.
92. Martin Ester, Hans-Peter Kriegel, Jorg Sander, and Xiaowei Xu. A density-based algorithm for discovering clusters in large spatial databases with noise. pages 226–231, 1996.
93. Julia Gunzenhäuser, Nicolas Olivier, Thomas Pengo, and Suliana Manley. Quantitative super-resolution imaging reveals protein stoichiometry and nanoscale morphology of assembling HIV-gag virions. *Nano Letters*, 12:4705–4710, 9 2012. ISSN 15306984. doi: 10.1021/nl3021076.
94. Sebastian Malkusch, Walter Muranyi, Barbara Müller, Hans Georg Kräusslich, and Mike Heilemann. Single-molecule coordinate-based analysis of the morphology of HIV-1 assembly sites with near-molecular spatial resolution. *Histochemistry and Cell Biology*, 139:173–179, 1 2013. ISSN 09486143. doi: 10.1007/s00418-012-1014-4.
95. Charlotte Floderer, Jean Baptiste Masson, Elise Boilley, Sonia Georgeault, Peggy Merida, Mohamed El Beheiry, Maxime Dahan, Philippe Roingeard, Jean Baptiste Sibarita, Cyril Favard, and Delphine Muriaux. Single molecule localisation microscopy reveals how HIV-1 Gag proteins sense membrane virus assembly sites in living host CD4 T cells. *Scientific Reports*, 8, 12 2018. ISSN 20452322. doi: 10.1038/s41598-018-34536-y.
96. Lucas von Chamier, Romain F. Laine, Johanna Jukkala, Christoph Spahn, Daniel Krentzel, Elias Nehme, Martina Lerche, Sara Hernández-Pérez, Pieta K. Mattila, Eleni Karinou, Séamus Holden, Ahmet Can Solak, Alexander Krull, Tim Oliver Buchholz, Martin L. Jones, Loïc A. Royer, Christophe Leterrier, Yoav Shechtman, Florian Jug, Mike Heilemann, Guillaume Jacquemet, and Ricardo Henriques. Democratising deep learning for microscopy with ZeroCostDL4Mic. *Nature Communications*, 12, 12 2021. ISSN 20411723. doi: 10.1038/s41467-021-22518-0.
97. Amin Zehtabian, Paul Markus Müller, Maximilian Goisser, Leon Obendorf, Lea Jänisch, Nadja Hümpfer, Jakob Rentsch, and Helge Ewers. Precise measurement of nanoscopic septin ring structures in deep learning-assisted quantitative superresolution microscopy. *bioRxiv*, 12 2021. doi: 10.1101/2021.12.28.474382.
98. Christophe Leterrier, Jean Potier, Ghislaine Caillol, Claire Debarnot, Fanny Rueda Boroni, and Bénédicte Dargent. Nanoscale architecture of the axon initial segment reveals an organized and robust scaffold. *Cell Reports*, 13:2781–2793, 12 2015. ISSN 22111247. doi: 10.1016/j.celrep.2015.11.051.
99. M. Heideman, D. Johnson, and C. Burrus. Gauss and the history of the fast fourier transform. *IEEE ASSP Magazine*, 1:14–21, 10 1984. ISSN 0740-7467. doi: 10.1109/MASSP.1984.1162257.
100. Hamidreza Heydarian, Florian Schueder, Maximilian T. Strauss, Ben van Werkhoven, Mohamadreza Fazel, Keith A. Lidke, Ralf Jungmann, Sjoerd Stallinga, and Bernd Rieger. Template-free 2D particle fusion in localization microscopy. *Nature Methods*, 15:781–784, 10 2018. ISSN 15487105. doi: 10.1038/s41592-018-0136-6.
101. J. A.G. Briggs. Structural organization of authentic, mature HIV-1 virions and cores. *The EMBO Journal*, 22:1707–1715, 4 2003. ISSN 14602075. doi: 10.1093/emboj/cdg143.
102. Denis Fortun, Paul Guichard, Virginie Hamel, Carlos Oscar S. Sorzano, Niccolo Banterle, Pierre Gonczy, and Michael Unser. Reconstruction from multiple particles for 3D isotropic resolution in fluorescence microscopy. *IEEE Transactions on Medical Imaging*, 37:1235–1246, 5 2018. ISSN 1558254X. doi: 10.1109/TMI.2018.2795464.
103. Swarup Dey, Chunhai Fan, Kurt V. Gothelf, Jiang Li, Chenxiang Lin, Longfei Liu, Na Liu, Minke A. D. Nijenhuis, Barbara Saccà, Friedrich C. Simmel, Hao Yan, and Pengfei Zhan. DNA origami. *Nature Reviews Methods Primers*, 1:13, 12 2021. ISSN 2662-8449. doi: 10.1038/s43586-020-00009-8.
104. Anette Eltner and Giulia Sofia. Structure from motion photogrammetric technique. *Developments in Earth Surface Processes*, 23:1–24, 1 2020. ISSN 09282025. doi: 10.1016/B978-0-444-64177-9.00001-1.
105. Hamidreza Heydarian, Maarten Joosten, Adrian Przybylski, Florian Schueder, Ralf Jungmann, Ben van Werkhoven, Jan Keller-Findeisen, Jonas Ries, Sjoerd Stallinga, Mark Bates, and Bernd Rieger. 3D particle averaging and detection of macromolecular symmetry in localization microscopy. *Nature Communications*, 12, 12 2021. ISSN 20411723. doi: 10.1038/s41467-021-22006-5.
106. Benjamin Blundell, Christian Sieben, Suliana Manley, Ed Rosten, QueeLim Ch'ng, and Susan Cox. 3D structure from 2D microscopy images using deep learning. *Frontiers in Bioinformatics*, 1, 10 2021. doi: 10.3389/fbinf.2021.740342.
107. Markus Mund, Aline Tschanz, Yu-Le Wu, Felix Frey, Johanna L Mehl, Marko Kaksonen, Ori Avinoam, Ulrich S Schwarz, and Jonas Ries. Superresolution microscopy reveals partial preassembly and subsequent bending of the clathrin coat during endocytosis. *bioRxiv*, 2022. doi: 10.1101/2021.10.12.463947.
108. Teun A. P. M. Huijben, Hamidreza Heydarian, Alexander Auer, Florian Schueder, Ralf Jungmann, Sjoerd Stallinga, and Bernd Rieger. Detecting structural heterogeneity in single-molecule localization microscopy data. *Nature Communications*, 12, 12 2021. ISSN 20411723. doi: 10.1038/s41467-021-24106-8.
109. Michael A. A. Cox and Trevor F. Cox. *Handbook of Data Visualization*. 2008.
110. A. K. Jain, M. N. Murty, and P. J. Flynn. Data clustering. *ACM Computing Surveys*, 31:264–323, 9 1999. ISSN 0360-0300. doi: 10.1145/331499.331504.
111. Alexander Auer, Maximilian T. Strauss, Sebastian Strauss, and Ralf Jungmann. Nan-

oTRON: A Picasso module for MLP-based classification of super-resolution data. *Bioinformatics*, 36:3620–3622, 6 2020. ISSN 14602059. doi: 10.1093/bioinformatics/btaa154.

112. Vilma Jimenez Sabinina, M. Julius Hossain, Jean Karim Hériché, Philipp Hoess, Bianca Nijmeijer, Shyamal Mosalaganti, Moritz Kueblbeck, Andrea Callegari, Anna Szymborska, Martin Beck, Jonas Ries, and Jan Ellenberg. Three-dimensional superresolution fluorescence microscopy maps the variable molecular architecture of the nuclear pore complex. *Molecular Biology of the Cell*, 32:1523–1533, 8 2021. ISSN 19394586. doi: 10.1091/mbc.E20-11-0728.